\documentclass[aps,prl,a4paper,showpacs,twocolumn,floatfix]{revtex4}
\usepackage[latin1]{inputenc}
\usepackage{amsfonts}
\usepackage{amssymb}
\usepackage{amsmath}
\usepackage{graphicx}
\usepackage{epsfig}
\usepackage{bm}
\usepackage{ulem}

\begin{document}

\title{Impedance-matched cavity quantum memory}
\date{\today}
\author{Mikael Afzelius$^1$ and Christoph Simon$^2$}
\affiliation{$^1$ Group of Applied Physics, University of
Geneva, CH-1211 Geneva 4, Switzerland\\ $^2$ Institute for
Quantum Information Science and Department of Physics and
Astronomy, University of Calgary, Calgary T2N 1N4, Alberta,
Canada}

\begin{abstract}
We consider an atomic frequency comb based quantum memory
inside an asymmetric optical cavity. In this configuration it is
possible to absorb the input light completely in a system
with an effective optical depth of one, provided that the
absorption per cavity round trip exactly matches the
transmission of the coupling mirror (``impedance
matching''). We show that the impedance matching results in a readout efficiency only limited by
irreversible atomic dephasing, whose effect can be made
very small in systems with large inhomogeneous broadening.
Our proposal opens up an attractive route towards quantum
memories with close to unit efficiency.
\end{abstract}

\maketitle

Quantum memories for photons
\cite{QAP,LvovskyNP,Hammerer2008,Tittel2010} are essential
elements for many applications in quantum information
processing, including quantum repeaters \cite{repeaters}
and linear-optics quantum computing
\cite{lin-opt-qc-review}. Most conceivable applications
require memories with storage and readout efficiencies that
are at or above the 90 \% level (and likely far above that
level for quantum computing). While quantum memory
experiments have progressed impressively over the last few years, efficiencies typically range from a few percent to a few tens of percent \cite{Chaneliere2005,Eisaman2005,Novikova2007,Choi2008,Riedmatten2008,Hosseini2009,Reim2009,Sabooni2009,Chaneliere2010,Usmani2010}. Only a few experiments have reached efficiencies above the 50\% level \cite{JSimon2007,hedges}, most notably a storage and readout efficiency close to 70\% has been achieved \cite{hedges} in a highly absorbing solid-state atomic ensemble using the gradient echo memory protocol \cite{GEM}.

It is usually thought that implementing memories in atomic
ensembles \cite{Hammerer2008} with efficiencies close to
unity will require optical depths much greater than one
\cite{Gorshkov2007a,AFC,hedges}. However, reaching high
optical depth is difficult in practice, in particular for
the most attractive solid-state systems, such as rare-earth
ion doped crystals \cite{Tittel2010}. Individual crystals
with realistic dimensions and doping levels often have very
limited optical depth. One exception is praseodymium-doped
Y$_2$SiO$_5$ crystals \cite{hedges}. But in order to fully
exploit the potential of other materials, having
considerably lower optical depth but otherwise interesting
coherence properties, it would be of great interest to find
a general method to overcome this crucial limitation.


Here we show that memories with unit efficiency can be
realized in a cavity-memory system with an optical depth of
one, by using the impedance matching condition. This
condition is attained \cite{Siegman} when the absorption, per
cavity round trip, is exactly matched to the transmission of
the coupling mirror of the (asymmetric) cavity. The result
is a complete absorption of the incoming light and we show
that the resulting memory readout efficiency reaches 100\%
for optical depths around 1. The use of impedance matching
had previously been suggested for quantum memories in
homogeneously broadened systems \cite{FYL}, however, the
results of Ref. \cite{Gorshkov2007a} later showed that in
such systems high effective optical depth is always
required for high efficiency, because the effect of
spontaneous decay cannot be ignored. In homogeneous systems
the efficiency roughly scales with $1-1/d$ \cite{Gorshkov2007a}, where $d$ is the optical depth.

Here we show that the situation is different in systems
with inhomogeneous broadening, for instance in solid-state approaches \cite{Tittel2010,AFC}. In such systems there is an
additional timescale given by the inverse of the
inhomogeneously broadened bandwidth. This can be much
shorter than the spontaneous decay time. As a consequence,
the effects of spontaneous decay can be negligible during
absorption and re-emission even for moderate optical depth.
Long storage times are nevertheless possible because the
inhomogeneous component of the dephasing can be made to be
reversible, e.g. by tailoring the spectral density in the
form of a frequency comb (AFC) \cite{AFC} or by using an externally controlled reversible inhomogeneous broadening (CRIB) \cite{Tittel2010,GEM}. As a consequence, the
principle of impedance matching can develop its full
potential in inhomogeneously broadened systems, as we will
now show in more detail.

\begin{figure}
    \centering
    \includegraphics[width=.3\textwidth]{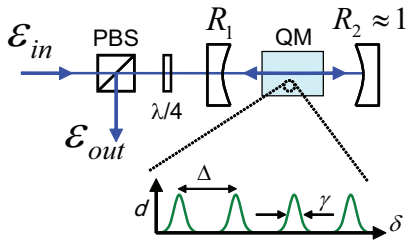}
    \caption{(Color online) We consider a quantum memory (QM) based
    on an atomic frequency comb which is placed in an asymmetric
    optical cavity with reflectivity $R_1<R_2 \approx 1$.
    The input and output fields ${\cal E}_{in}$ and ${\cal E}_{out}$ are
    separated by a quarter-wave plate ($\lambda/4$) and
    a polarization beam splitter (PBS). If the QM strongly absorbs only
    a particular linear polarization mode, one can also use a
    Faraday rotator and a half-wave plate as in Ref. \cite{Usmani2010}. The
    atomic comb memory is based on an inhomogeneously broadened transition,
    where
    the absorption depth $d$ is shaped into a comb structure as function of
    detuning $\delta$ with periodicity $\Delta$ and tooth width $\gamma$.
    The interaction between the atomic comb structure and a
    incoming light pulse leads to a coherent re-emission at $t=2\pi/\Delta$.
    Longer storage times can be achieved by using additional ground state
    levels \cite{AFC,Afzelius2010}.}
    \label{fig_setup}
\end{figure}

Let us start by considering the absorption of light by an
inhomogeneously broadened atomic ensemble in a one-sided cavity, see Fig. \ref{fig_setup}. The readout step will be treated later, for the case of an AFC-based control of the inhomogeneous dephasing. The dynamical equation for the cavity field ${\cal E}$ is
\begin{equation}
\dot{\cal{E}}=-\kappa {\cal E} +\sqrt{2\kappa}{\cal
E}_{in}+ i \tilde{\cal{P}} \int d \omega n(\omega)
\sigma_{\omega}, \label{cavity}
\end{equation}
where $\kappa$ is the cavity decay rate, $\tilde{{\cal P}}$
is proportional to the dipole moment \cite{AFC}, $\omega$
is the detuning, $n(\omega)$ is the inhomogeneous atomic
spectral distribution and $\sigma_{\omega}$ is the atomic
polarization at detuning $\omega$. The equation for the
atomic polarization is
\begin{equation}
\dot{\sigma_{\omega}}=-i\omega \sigma_{\omega}-\gamma_h
\sigma_{\omega}+i{\cal P} {\cal E}, \label{polarization}
\end{equation}
where $\gamma_h$ is the homogeneous linewidth and ${\cal
P}$ is the dipole moment. Finally the input-output relation
for the cavity is
\begin{equation}
{\cal E}_{out}=-{\cal E}_{in}+\sqrt{2\kappa} {\cal E},
\label{inout}
\end{equation}
which is valid for relatively high cavity finesse. (We will
drop this simplifying assumption later on.)

Putting the solution of Eq. (\ref{polarization}) into Eq.
(\ref{cavity}) gives
\begin{eqnarray}
\dot{\cal{E}}(t)=-\kappa {\cal E}(t) +\sqrt{2\kappa}{\cal
E}_{in}(t)- \nonumber\\
{\cal P}\tilde{\cal{P}} \int
\limits_{-\infty}^t dt' \tilde{n}(t-t') e^{-\gamma_h(t-t')}
{\cal E}(t'),
\end{eqnarray}
where $\tilde{n}(t)$ is the Fourier transform of
$n(\omega)$. If $\gamma_i \gg \gamma_h$, where $\gamma_i$
is the width of the inhomogeneous distribution $n(\omega)$,
then the exponential containing $\gamma_h$ can be ignored
over the relevant timescales. If moreover $\gamma_i$ is
significantly larger than the bandwidth of the input light,
then $\tilde{n}(t-t')$ can be approximated as
$\frac{N}{\gamma_i}\delta(t-t')$ (for times around
$t=0$, i.e. when the absorption happens, cf. below for much
later times), where $N=\int d\omega n(\omega)$ is the
total number of atoms, yielding
\begin{equation}
\dot{\cal{E}}=-\kappa {\cal E}+\sqrt{2\kappa} {\cal
E}_{in}-\Gamma {\cal E}, \label{Gamma}
\end{equation}
where $\Gamma=\frac{N{\cal P}\tilde{\cal{P}}}{\gamma_i}$
emerges as the absorption rate of the cavity field by the
atomic ensemble.

Under conditions where the input field varies much more
slowly than the cavity lifetime, i.e. when the input
spectrum is in resonance with the cavity, one can now
adiabatically eliminate the cavity mode (i.e. set
$\dot{{\cal E}}=0$), which gives
\begin{equation}
{\cal E}=\frac{\sqrt{2\kappa}}{\kappa+\Gamma}{\cal E}_{in}.
\end{equation}
Plugging this into Eq. (\ref{inout}) results in
\begin{equation}
{\cal E}_{out}=\frac{\kappa-\Gamma}{\kappa+\Gamma} {\cal
E}_{in}. \label{kappaGamma}
\end{equation}
Total absorption, corresponding to ${\cal E}_{out}=0$, can
thus be achieved for $\kappa=\Gamma$, which is the
impedance matching condition in our case. The intuitive
explanation is that in this situation the absorption losses
have exactly the same effect as a second identical mirror
would. To the input field the cavity-memory system
therefore looks exactly like a symmetric Fabry-Perot
cavity, leading to zero reflection on resonance \cite{Siegman}. The ratio
$\Gamma/\kappa$ is exactly the effective optical
depth, or the cooperativity $C$ in the notation of Ref.
\cite{Gorshkov2007a}. Perfect absorption is thus achieved
for an optical depth of one, a very moderate value. Our
results are nevertheless consistent with those of
Ref. \cite{Gorshkov2007a} in the sense that if all $N$
atoms were concentrated into the homogeneous linewidth
$\gamma_h$ rather than the inhomogeneous linewidth
$\gamma_i$, the resulting cooperativity would be very
large, given our assumption that $\gamma_i \gg \gamma_h$.
However, fortunately there is no need for all the $N$ atoms
to actually have the same frequency in the quantum memory
schemes based on control of inhomogeneous dephasing.

In the context of quantum memories it is crucial to also
obtain an efficient readout of the stored excitation. Here we will limit our analysis to the case of an AFC-based \cite{AFC} quantum memory. We only briefly remind the reader of the essential features, for details we refer to Ref. \cite{AFC}. The inhomogeneous absorption is shaped into a comb structure, by optical pumping techniques, having periodicity $\Delta$ and peak width $\gamma$ (see Fig. \ref{fig_setup}). The interaction between an incoming light pulse in resonance with the comb results in a coherent re-emission after a time $t=2\pi/\Delta$, due to a periodic rephasing of the atomic coherence (we assume that the input spectrum is larger than $\Delta$). Note that freely controllable
storage times far beyond $2\pi/\Delta$ can be achieved by using an additional ground state level \cite{AFC}, as recently also shown experimentally \cite{Afzelius2010}.

In the case of a high AFC comb finesse $F_A=\Delta/\gamma$, the efficiency of this echo-type emission can be very large for large optical depths \cite{AFC}. For a forward readout configuration it is limited to 54\% due to re-absorption in the sample, while for a backward readout it can reach 100\% due to an interference effect that is well-understood \cite{AFC}. We will show below that in our proposed cavity arrangement, the efficiency can reach 100\% for a much lower optical depth, also without having to resort to the backward recall procedure \cite{AFC}.

We thus assume that $n(\Delta)$ has the shape of a comb, as in Fig. \ref{fig_setup}. As a consequence $\tilde{n}(t)$ has peaks not only at $t=0$ (as we used before), but also at integer multiples of
$2\pi/\Delta$. Following Ref. \cite{AFC} one can derive the following
equation for the cavity field around the first rephasing at $t=2\pi/\Delta$,
\begin{eqnarray}
\dot{\cal{E}}(t)=-\kappa {\cal E}(t) - \frac{{\cal
P}\tilde{\cal{P}}}{\sqrt{2\kappa}} \int \limits_{-\infty}^0
dt' \tilde{n}(t-t') {\cal E}_{in}(t')\nonumber\\
- {\cal P}\tilde{\cal{P}} \int \limits_{0}^t dt'
\tilde{n}(t-t') {\cal E}(t'),
\end{eqnarray}
similar to Eq. (A15) in Ref. \cite{AFC}. Using similar
arguments as for the absorption, this reduces to
\begin{equation}
\dot{\cal{E}}(t)=-\kappa {\cal E}(t)-2 \frac{\Gamma
\sqrt{\eta_F}}{\sqrt{2\kappa}} {\cal
E}_{in}(t-\frac{2\pi}{\Delta})-\Gamma {\cal E}(t),
\end{equation}
which is analogous to Eq. (A16) in Ref. \cite{AFC}. Here
$\eta_F$ describes the reduction in efficiency due to the
fact that the individual teeth in the frequency comb have
finite width, which leads to irreversible atomic dephasing. In the case of Gaussian peaks \cite{AFC} one finds $\eta_F\approx e^{-7/F_A^2}$. This should not be
confused with the cavity finesse ${\cal F_C}=\frac{\pi (R_1
R_2)^{\frac{1}{4}}}{1-\sqrt{R_1 R_2}}$.

Adiabatically eliminating the cavity mode as before and
using the fact that there is no input field at
$t=2\pi/\Delta$ we find
\begin{equation}
{\cal E}_{out}(t)=-\frac{2 \Gamma
\sqrt{\eta_F}}{\kappa+\Gamma} {\cal
E}_{in}(t-\frac{2\pi}{\Delta})=-\sqrt{\eta_F} {\cal
E}_{in}(t-\frac{2\pi}{\Delta}),
\end{equation}
where the last equality holds under the impedance matching
condition ($\kappa=\Gamma$). One sees that the readout efficiency is only
limited by the finesse of the atomic frequency comb, which without the cavity would correspond to an infinitely high optical depth $d$ \cite{AFC}.

The above treatment applies to the regime where $R_1
\approx 1$ and $R_2=1$. More precise results for a general
asymmetric cavity can be obtained in the following way. For
the absorption it is sufficient to include absorption
factors into the usual ``sum over all roundtrips''
treatment of a Fabry-Perot cavity. This yields
\begin{equation}
{\cal E}_{out}={\cal E}_{in}
\frac{-\sqrt{R_1}+\sqrt{R_2}e^{-\tilde{d}}}{1-\sqrt{R_1
R_2} e^{-\tilde{d}}}
\end{equation}
on resonance, where $\tilde{d}$ is the optical depth of the
crystal inside the cavity (averaged over the frequency
comb, cf. Ref. \cite{AFC}). One sees that perfect
absorption is still achievable provided that
$\sqrt{R_1}=\sqrt{R_2} e^{-\tilde{d}}$, which is the impedance condition \cite{Siegman}.

\begin{figure}[hbt]
    \centering
    \includegraphics[width=.45\textwidth]{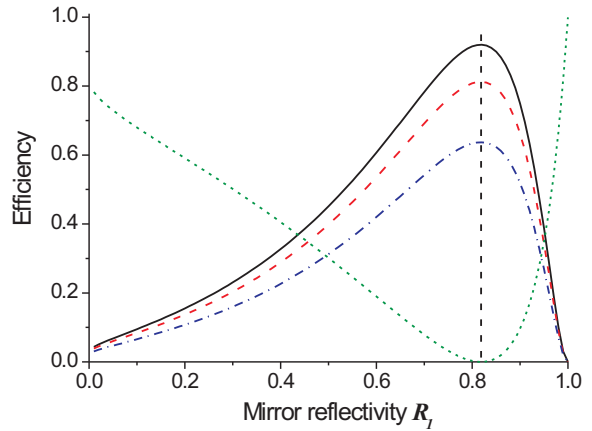}
    \caption{(Color online) We here show the efficiency of an AFC-cavity
    quantum memory in an asymmetric cavity ($R_2$=0.999) as a function of
    the input mirror reflectivity $R_1$, based on Eq. \ref{exacteta}.
    We show the result for different comb finesse $F_A$=10 (solid line),
    $F_A$=6 (dashed line) and $F_A$=4 (dashed-dotted line). The single-pass
    effective absorption depth was set to $\tilde{d}$=0.1, which in a memory
    without cavity would bound the efficiency to $\sim$1\% (by use of Eq.
    \ref{singlepass_eta}). In the figure one clearly observes the great
    enhancement of memory efficiency using an impedance-matched cavity,
    i.e. at $R_1=\exp(-2\tilde{d})\approx0.82$, reaching $\eta\sim$92\%
    for $F_A$=10. At this point the efficiency is only limited by irreversible
    atomic dephasing due to the finite comb finesse $F_A$. We also plot the
    reflectivity of the combined AFC-cavity system (dotted line), showing
    the complete absorption of light at the optimal point.}
    \label{fig_eff_vs_R1}
\end{figure}

A similar treatment is possible for the memory readout.
From Ref. \cite{AFC} it is known that the readout
efficiency can be obtained via a ``sum over all
amplitudes'' approach. For example, for forward readout the
relevant efficiency factor is given by Eq. (A19) of Ref.
\cite{AFC},
\begin{equation}
\int \limits_0^L dz e^{-\tilde{\alpha} z/2}\tilde{\alpha}
e^{-\tilde{\alpha} (L-z)/2}=\tilde{\alpha}L
e^{-\tilde{\alpha}L/2},
\label{singlepass_eta}
\end{equation}
where $L$ is the length of the crystal, $\tilde{\alpha}$ is
the absorption coefficient ($\tilde{\alpha}L=\tilde{d}$),
and one integrates over all possible points of absorption
$z$. The first factor under the integral corresponds to the
amplitude for the photon to be transmitted to the point
$z$, the second factor can be interpreted as the amplitude
for absorption and re-emission (in $z$), and the third
factor is the amplitude to be transmitted from $z$ to the
end of the crystal after re-emission. This can be
generalized for a Fabry-Perot cavity, taking into account
the fact that the photon can do an arbitrary number of
round trips in the cavity before absorption and after
re-emission. The result is
\begin{eqnarray}
2\int dz
\frac{\sqrt{T_1}e^{-\tilde{\alpha}z/2}}{1-\sqrt{R_1 R_2}
e^{-\tilde{d}}} \tilde{\alpha}
\frac{e^{-\tilde{\alpha}(L-z)/2}
e^{-\tilde{\alpha}L/2}\sqrt{T_1 R_2}}{1-\sqrt{R_1 R_2}
e^{-\tilde{d}}},
\end{eqnarray}
where $T_1=1-R_1$ is the transmission of the first mirror.
Again the first factor under the integral corresponds to
propagation before absorption, the second factor is the
absorption and re-emission amplitude, and the third factor
is for propagation after re-emission. The factor of 2 in
front of the integral stems from the fact that the photon
can be absorbed while propagating either in forward or in
backward direction. Note that inside the cavity there is no
change of direction upon re-emission. (Of course, the
output field of the asymmetric cavity propagates
predominantly in the opposite direction to the input field,
but this is an automatic consequence of the interference
between all the possible paths.) Simplifying the above
expression, and multiplying by $\sqrt{\eta_F}$ to take into
account the irreversible component of the atomic dephasing,
one obtains the following expression for the square root of
the total memory efficiency $\eta$ (as is customary for
quantum memories, we define efficiencies with respect to
intensities, not amplitudes),
\begin{equation}
\sqrt{\eta}=\frac{2 \tilde{d} e^{-\tilde{d}} T_1 \sqrt{R_2}
\sqrt{\eta_F}}{(1-\sqrt{R_1 R_2} e^{-\tilde{d}})^2}.
\label{exacteta}
\end{equation}
Our previous results correspond to the limit
$\sqrt{R_1}=1-\epsilon$ with $\epsilon \ll 1$, $R_2=1$,
$\tilde{d} \ll 1$. In this case Eq. (\ref{exacteta})
becomes
\begin{equation}
\sqrt{\eta}=\frac{2 \tilde{d}
\sqrt{\eta_F}}{\epsilon+\tilde{d}},
\end{equation}
so that we recover our previous result ($\eta=\eta_F$)
under the impedance matching condition, which is now
expressed as $\epsilon=\tilde{d}$.

The total memory efficiency $\eta$ (which includes
absorption and re-emission) is shown in Fig.
\ref{fig_eff_vs_R1} as a function of input mirror
reflectivity $R_1$. Clearly one can achieve very high
efficiency for low reflectivities (in the context of
optical cavities) and for very reasonable optical depths.
For example, a memory with a peak optical depth $d$=1
and AFC finesse $F_A$=10, such that $\tilde{d}=0.1$, has
an efficiency of only 1\% without cavity, but can be
boosted to 92\% efficiency by an impedance-matched cavity
of finesse ${\cal F_C}\approx31$.

An impedance-matched cavity memory can be operated for a large variety of conditions. Equation (\ref{exacteta}) allows us to find the best working conditions for a particular situation. There are some assumptions, however, that must be fulfilled. The quantum memory bandwidth must be significantly smaller than the width of the optical cavity in order to fulfill the resonance condition used above. As an example, if we assume a cavity length $L$=1 cm (reasonable for typical crystal dimensions) the cavity width would be $\approx$480 MHz for the example above. We have also assumed that the cavity has no losses. In general the losses must clearly be significantly lower than the memory absorption probability (per single pass). The effect of losses can be evaluated, however, by changing the reflectivity of the second mirror $R_2$, thus introducing a loss to the environment. For the example above, $R_2$=0.99 instead of $R_2$=0.999 would reduce the efficiency to 84\%. Practically a good AR-coating on the crystal should keep the losses low enough.

In conclusion, we have shown that impedance matching to an optical cavity allows the implementation of highly efficient
quantum memories for an effective optical depth of only one. Our proposal should make it much easier for experiments to reach the truly high efficiency regime.

The authors acknowledge useful discussions with H. de Riedmatten, N. Sangouard, N. Gisin and J.-L. Le Gou\"{e}t. This work was supported by the Swiss NCCR Quantum Photonics, the EU projects Quessence and QuRep.

{\it Note added.} When this work was completed, we became aware of a recent related proposal \cite{moiseev2010}.


\end{document}